%% file: bprv1.tex
\documentclass[12pt,titlepage,twoside,rotate,epsf,graphics]{article}
\usepackage{english,a4,makeidx,latexsym,epsfig}

\newsavebox{\LSIM}
\sbox{\LSIM}{\raisebox{-1ex}{$\ \stackrel{\textstyle<}{\sim}\ $}}
\newcommand{\lsim}{\usebox{\LSIM}}
\newsavebox{\GSIM}
\sbox{\GSIM}{\raisebox{-1ex}{$\ \stackrel{\textstyle>}{\sim}\ $}}
\newcommand{\gsim}{\usebox{\GSIM}}

\begin{document}
\begin{titlepage}
\begin{flushright}
HD-THEP-99-25\\
\end{flushright}
\begin{centering}
\vfill
 
{\bf \Large Bosonic Preheating in Left-Right-Symmetric SUSY GUTs}
\vspace{1cm}
 
R. Micha\footnote{r.micha@thphys.uni-heidelberg.de}, 
M.G. Schmidt\footnote{m.g.schmidt@thphys.uni-heidelberg.de} \\
 
\vspace{1cm} {\em 
Institut f\"ur Theoretische Physik, 
Philosophenweg 16, 
D-69120 Heidelberg, Germany}\\

\vspace{2cm}
 
{\bf Abstract}
 
\vspace{0.5cm}
We investigate the possibility of a bosonic preheating in the simplest model of supersymmetric Hybridinflation (F-term inflation), which was considered first by Dvali et al. Here the inflationary superpotential is of the O'Raifertaigh-Witten type. The end of inflation is related to a non-thermal phase transition, which in the context of left-right symmetric models lowers the rank of the gauge group. Using the homogeneous classical field ansatz for the appearing condensates, our results indicate that the parametric creation of bosonic particles does not occure in the model under consideration.
\end{centering}
 
\vspace{0.3cm}\noindent

\vfill \vfill
\noindent
 
\end{titlepage}
 

\section{Introduction}
The recent success in detecting neutrino masses has caused a renaissance of
the idea of
grand unification.
In left-right-symmetric models, such  as $SU(3)_c\times SU(2)_L\times SU(2)_R\times U(1)_{(B-L)}$ or $SO(10)$, massive Majorana neutrinos arise
naturally. As the difference of baryon- and lepton number is a gauged symmetry
in these models, there could exist interesting mechanisms for the creation of the
corresponding asymmetries. On the other hand, supersymmetry
not only provides the right mass scale for the heavy neutrinos,
but also gives the possibility of a inflationary potential in the
GUT-Higgs-Sektor. Then it is possible to construct a cosmological
model based on a consistent supersymmetric grand unified theory, which can be judged by
both, its cosmological and its particle theoretical features. In this paper we
will concentrate on an O'Raifertaigh-Witten model, which by the autors of \cite{dvali} was shown
to be a realisation of Linde's Hybrid Inflation scenario \cite{lin3}. The
superpotential can serve  as a part of the Higgs sector of a left-right-symmetric model with global or local supersymmetry. The phase of inflation and the
formation of density fluctuation in this model have been studied in \cite{dvali,lin2}. Since
a successful cosmological model needs a very effective mechanism for particle production after
inflation, here we  investigate the possibility of a bosonic
preheating in the inflationary potential under consideration.
As preheating possibly allows for the creation of superheavy particles, it is
a very interesting scenario not only in the context of lepto- and baryogenesis
but also for the creation of dark matter.
Focusing on qualitative insights rather than quantitative accuracy,
we tried to work with a minimum of numerical expense. Our investigation was
inspired by those in \cite{bellido2,bellido1}, where preheating in the original
version of Hybrid Inflation was
studied. Our numerical results, which originally were presented in the context of an investigation of inflationary supersymmetric $SO(10)$-models in \cite{da}, are very similiar to those in a recent paper, based on a version of a NMSSM \cite{susypr}. But, as shall be seen, we disagree with the physical
interpretation of the results given there. In contrast to \cite{susypr}, we do not see any evidence for bosonic preheating in supersymmetric hybrid inflation.

\section{Supersymmetric Hybridinflation: the Basic Scenario}
\label{sec:2}
\setcounter{figure}{0}
\setcounter{equation}{0}
A common way to achieve an inflationary scenario from a supersymmetric
theory, is to use the simplest  O'Raifertaigh-Witten model for
spontaneous symmetry breakdown \cite{dvali}. It is  given by the superpotential
\begin{equation}\label{RWPot}
W=X(\kappa {\bar C}C-\mu^2),
\end{equation}
where the couplings are protected by either a continouus or a discrete
$R$-symmetry, which could descend from a string theory. The superfield
$C=(\hat a_C,\hat \psi_C)$ is a representation of the Lie-algebra $A_G$ of the Lie-group G, and $\bar C=(\hat a_{\bar C},\hat
\psi_{\bar C})$ is the conjugate representation. Here the $\hat a$'s represent the
scalar component while the $\hat \psi$'s are the spinors. The Superfield $X=(\hat a_X,\hat
\psi_X)$ is an $A_G$-Singlet. The mass-scale $\mu$ could be caused by a string compactification. In the case, that $G$ represents a gauge
symmetry we have the tree-level scalar potential
\begin{eqnarray}
  V^{(0)}&=&|F_X|^2+|F_{C}|^2+|F_{\bar C}|^2+\frac{1}{2}\sum\limits_{r=1}^{dim(G)}|D_r|^2\\
&=&\kappa^2|a_Xa_{\bar C}|^2+\kappa^2|a_Xa_{C}|^2+|\kappa a_{\bar
  C}a_{C}-\mu^2|^2 +\frac{1}{2}\sum\limits_{r=1}^{dim(G)}|D_r|^2\nonumber.
\end{eqnarray}
Here $a_i$ represents the vacuum expectation value (vev) of the quantum field
$\hat a_i$. Then $V^{(0)}$ is minimized for $\arg{a_{\bar C}}+\arg{a_{ C}}=0$ and it is
independent of $\arg{a_{\bar C}}+\arg{a_{ X}}$ and
$\arg{a_{C}}+\arg{a_{X}}$. Supersymmetry requires the $D$-terms to vanish
identically for each single generator of the group $G$. If $G$ represents a
product group including an $U(1)$-factor, this means that $|a_C|=|a_{\bar C}|$.

In the following we will be interested only in the direction of the
$A_{G}$-multiplets $a_{C}$ and $a_{\bar C}$, which aquire a G breaking
vev. Therefore they will be denoted by the same names.
Now, concentrating on the $D=0$-direction, there exists a certain
$R$-transformation \cite{laz} which brings the scalar components $a_i$ to the
real axis. The corresponding canonically normalized scalarfields $\phi$,
$\sigma$ and couplings $\lambda$, $g$ are given by \cite{laz}:
\begin{eqnarray}
\sigma&:=& {2}a_C={2}a_{\bar C}\\
\phi &:=& {\sqrt{2}}a_X\\
\lambda&:=&\frac{\kappa^2}{4}\\
g&:=&\frac{\kappa}{\sqrt{2}}\\
M&:=&\sqrt{\kappa}\mu, 
\end{eqnarray}
where typically $g^2=2\lambda$.
Using this conventions the scalar effective potential in terms of the fields
$\phi$ and $\sigma$ reads 
\begin{equation}\label{Pot}
V(\phi ,\sigma)=\frac{1}{4\lambda}\left( M^2-\lambda \sigma^2   \right) ^2+\frac{1}{2}g^2\phi^2\sigma^2+V^{(1)}(\phi),
\end{equation}
which resembles very much Linde's original model of Hybrid Inflation \cite{lin3}.
Here $V^{(1)}(\phi,\sigma)$ represents the loop corrections to the tree level
potential, which vanish for the supersymmetric case. For $\sigma$=$0$, $\phi>M/g=:\phi_c$ the gauge symmetry remains
conserved while supersymmetry is broken in the sector of the gauge singlet $X$
inducing loop-corrections $V^{(1)}(\phi,\sigma)$. Including 1-loop-corrections
the effective
potential in this regime is given by \cite{dvali}
\begin{eqnarray}
V(\phi >\phi_c)&=& \frac{M^4}{4\lambda}\left( 1+ \frac{\lambda}{4\pi^2} \left[ 
  \ln{ \frac{2\lambda\phi^2}{M^2}} + \left( \frac{2\lambda\phi^2}{M^2}-1\right)^2\ln{\left(
  1-\frac{M^2}{2\lambda\phi^2} \right) }+{}\right. \right.     \nonumber \\ 
   &&{} + \left. \left. \left( \frac{2\lambda\phi^2}{M^2}+1\right)^2 \ln{\left(
  1+\frac{M^2}{2\lambda\phi^2} \right) }+\ln \frac{M^2}{\Lambda^2}
  \right]\right),
\end{eqnarray}
where $\Lambda$ is a renormalization scale. Since soft susy breaking terms
lead to scalar masses of
${\cal O}$(TeV), such terms are neglegible compared to the GUT-scale mass parameter $M$.
For a large range of the parameters $\kappa$ and $\mu$ the potential above satisfies the
Slow Roll conditions for inflation. Here
inflation is caused by the ``cosmological constant'' $\frac{M^4}{4\lambda}$. The
scalar gauge singlet is the only degree of freedom, which then has a nonvanishing
vev $\phi$. This vev very soon is dominated by the zero momentum Fourier
mode and it forms a  condensate with a very small extension  in
momentum space. Referring to \cite{lin1} it can
be approximated by a homogeneous classical scalar field. For this reason we
will call $\phi$ the inflaton and identify it with the zero momentum mode. The slow roll regime of the singlet is a
typical aspect of the superpotential (\ref{RWPot}) and has become popular as
effect of ``a sliding field'' outside the context of inflation \cite{schmidt1}. 

The end of
inflation is connected to the end of the gauge symmetric phase: as soon as $\phi$ reaches  the critical point $\phi_c$  the
effective mass $m_{\sigma}^2 = -M^2+g^2\phi^2$ of the Higgs-field
vanishes. The classical equations of motions alone cannot tell us something about a 
phase transition, since  the derivative $V(\phi_c,\sigma=0),_{\sigma}$ vanishes
identically and the  vev of $\hat \sigma $ remains zero. But, following the argumentation of Garcia-Bellido and Linde \cite{bellido2}, as $\phi$ slides
towards zero, quantum fluctuations around this vev will get tachyonic
masses stimulating  an  exponential growth of
modes, whose momenta are smaller than
the effective mass, $k<|m_{\sigma}|$, where $k$ means the comoving momentum. The modes with $k>|m_\sigma|$ will not grow at
all. The result is a spontaneous breakdown of the gauge symmetry caused by the 
inhomogenous distribution of the field $\hat \sigma$ with $ \langle\hat\sigma\rangle = 0$. Inflation ends typically near this phase
transition. 

We now want to understand, if the behaviour of the Higgs field $\hat \sigma$
can be approximated by an homogeneous classical field. 
As pointed out in \cite{bellido2} the distribution of the Higgs field would be
homogeneous on scales $l \sim |m_{\sigma}|^{-1}$ or even somewhat greater, if $m_{\sigma}$ was constant in
time. However,  we will see soon that  this mass  is oscillatory. This changes
the range and the mechanism of amplification of the ``tachyonic modes''. Due
to  this fact, the distribution $\sigma_k(t)$ in momentum space will be even
more localized around the zero momentum mode, such that the spatial
distribution will be homogeneous even on a scale $l$ which is somewhat greater than $ |m_{\sigma}|^{-1}$.
At least at such scales $\hat \sigma(t,\vec x)$ also can be effectively described by a classical
homogenous  scalar field $\sigma(t)$, i.e. as the zero momentum
mode of such a classical field, which rolls down from the critical
point to the supersymmetric minimum, where $\sigma =M/\sqrt{\lambda}$. We will
call $\sigma(t)$ the Higgs-condensate. The range of validity for this ansatz
will become clearer during our numerical investigation.

As pointed out in \cite{lin2} local supersymmetry modifies this scenario
because of the  non-renormalizable terms, whose influence on the inflation depends on the value of
the coupling constant $\kappa$. But in  most of these cases the end of
inflation is dominated by the renormalizable terms in the potential. Thus,
non-renormalizable couplings will not contribute after the phase transition and will be ignored in the
following investigation.

\section{Evolution of the Background }
\label{sec:3}
\setcounter{figure}{0}
\setcounter{equation}{0}
After the phase transition both of the fields $\hat\phi(t, \vec x)$
and $\hat\sigma(t, \vec x)$ form condensates, which essentially can be described by
the two  homogeneous classical fields $\phi(t)$ and
$\sigma(t)$. Free Higgs or singlet particles will be  identified with
quantum fluctuations around these condensates.
In this section we concentrate on the evolution of the background and 
we neglect the effect of the quantum fluctuations.  
For reasons of simplification we use a ``M-rescaling'' to natural variables:
 \begin{eqnarray}
t &\rightarrow & y:=Mt\nonumber\\
\vec x  &\rightarrow & \vec \xi:=M\vec x\nonumber\\
\vec k  &\rightarrow & \vec K:=\frac{\vec k}{M}\\
\phi&\rightarrow & f:=\frac{\phi g}{M}=\frac{\phi}{\phi_c}\nonumber\\
\sigma&\rightarrow & s:=\frac{\sigma \sqrt{\lambda}}{M}=\frac{\sigma}{\sigma_c}.\nonumber
\end{eqnarray}
Here the comoving momenta $\vec K$ and the scale factor $a(y)$ are normalized in such a way that $a(y_c)=1$ is fullfilled, where $y_c$ means the M-rescaled time at the phase transition.
\begin{figure}
\begin{minipage}[t]{6.5cm}
\vspace*{0cm}
\begin{picture}(0,0)
\put(160,-125){$\frac{\phi}{\phi_c}$}
\put(160,-70){$\frac{H}{H_0}$}
\put(160,-15){$\frac{\sigma}{\sigma_0}$}
\put(10,0){\epsfysize=6cm\rotatebox{270}{\epsffile{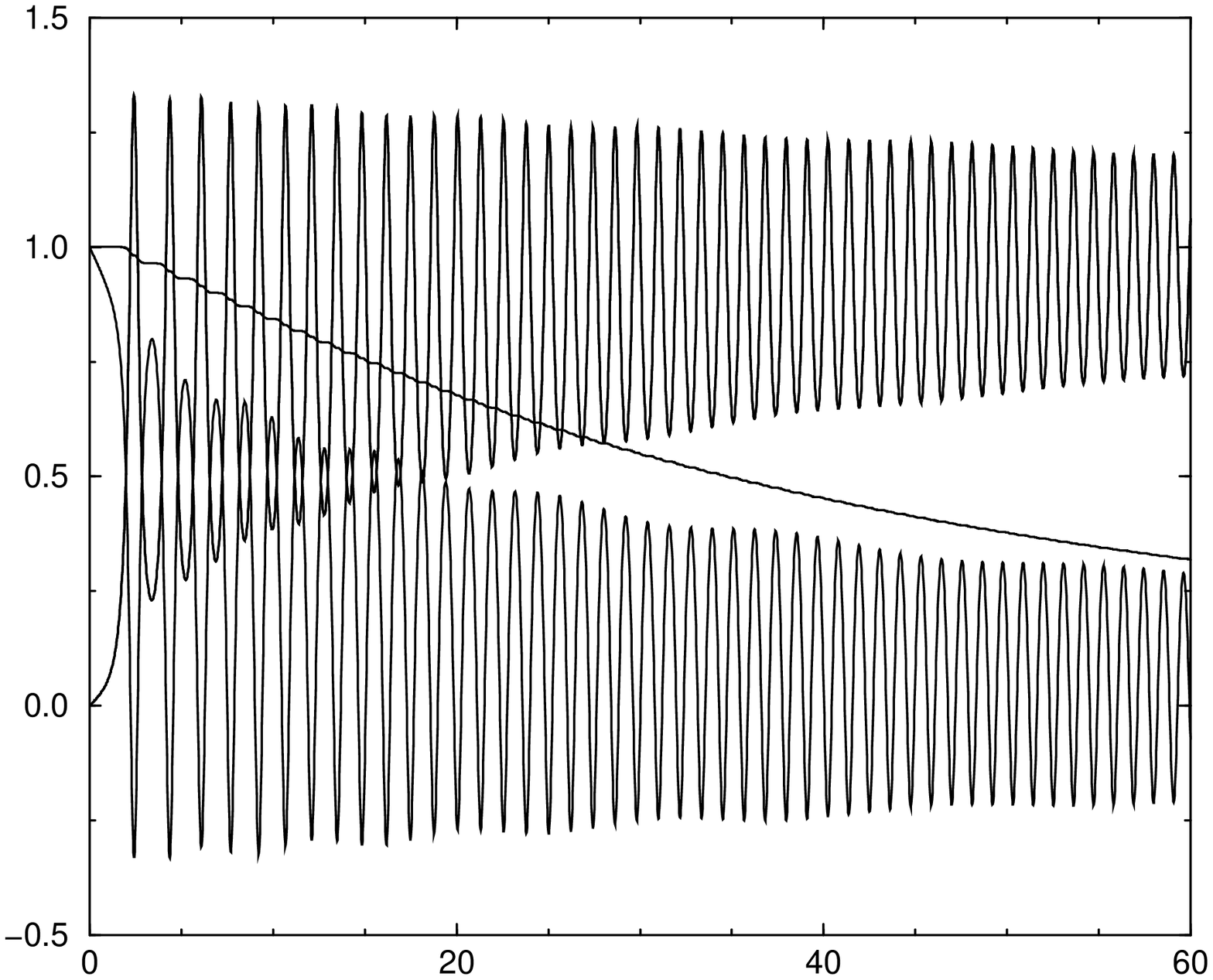}}}
\put(87,-150){$\frac{\sqrt{2}Mt}{2\pi}$}
\end{picture}
\vspace*{5cm}
\caption{Evolution of the background}\label{Linde0}
\end{minipage}\hfill
\begin{minipage}[t]{6.5cm}
\begin{picture}(0,0)
\put(0,-70){$\frac{\phi}{\phi_c}$}
\put(10,0){\epsfysize=6cm\rotatebox{270}{\epsffile{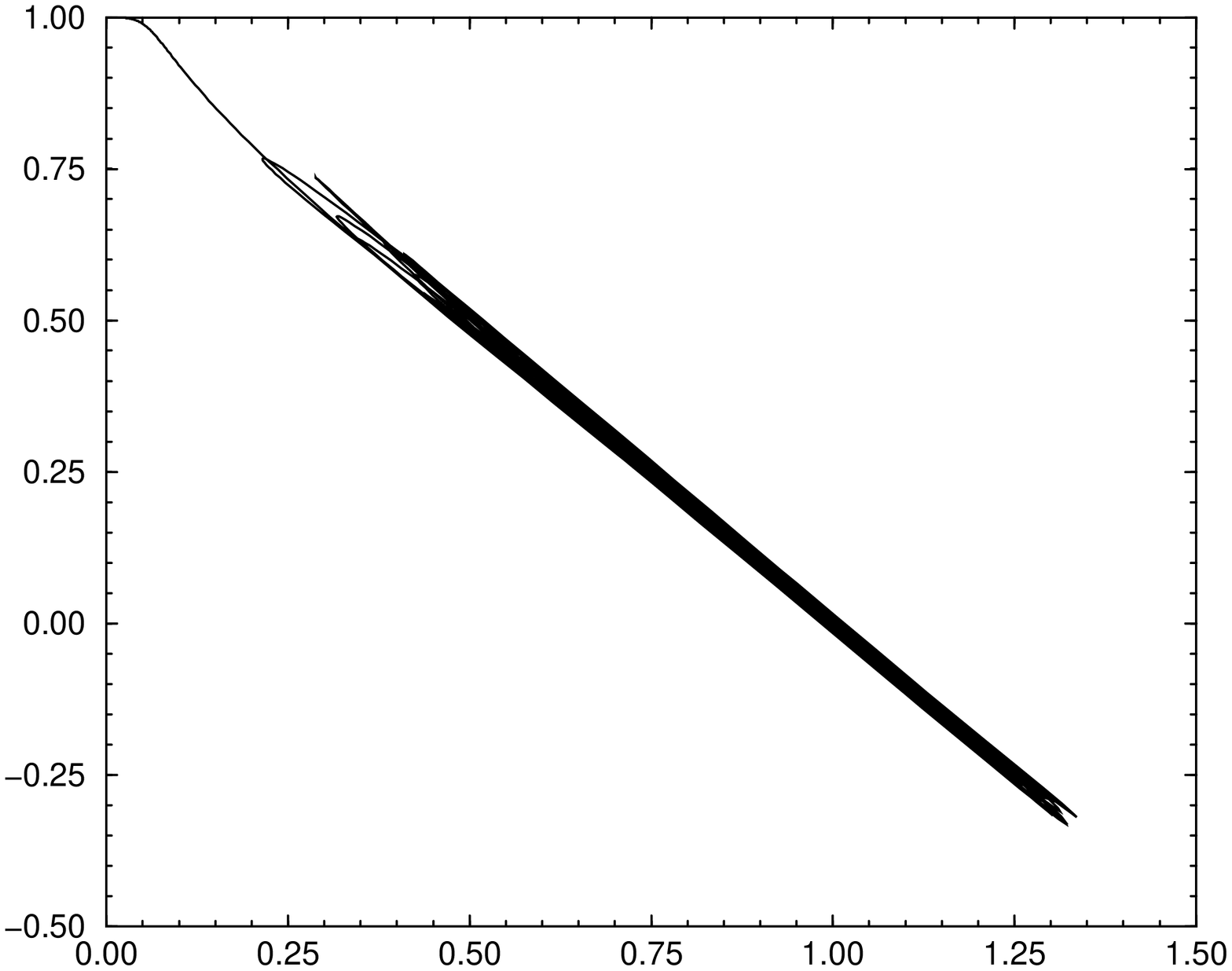}}}
\put(95,-150){$\frac{\sigma}{\sigma_0}$}
\end{picture}
\label{fig3}
\vspace*{5cm}
\caption{Oscillations in the $\phi$-$\sigma$-plane }\label{chaos}
\end{minipage}
\end{figure}

Neglecting the sub-dominant loop corrections, which certainly  will appear, as long as the
supersymmetric minimum is not reached, the equations of motion after the phase transition read:
\begin{eqnarray}
&\left( \frac{d^2}{dy^2} +3h(y)\frac{d}{dy}+2s^2(y) \right)  f (y)=0&\label{nphiBgl}\\
&\left(\frac{d^2}{dy^2} +3h(y)\frac{d}{dy}+\left(-1+f^2(y)+s^2(y) \right) \right)  s (y)=0&\label{nsigmaBgl}
\end{eqnarray}
where  $h$ is the M-rescaled Hubble Parameter, $h:=(\frac{d}{dy}a)/a$, with
\begin{equation}
h^2(y)=\frac{2\pi}{3}\frac{M^2}{\lambda M_{pl}^2}\left[2 {\dot f}^2(y)+{\dot s}^2(y)+ (1-s^2(y) )^2+2f^2(y)s^2(y) \right].\label{nHubble}
\end{equation}
Important qualitative insights then are possible without any detailed calculation: Both of the
classical fields, $f=\frac{\phi}{\phi_c}$ and $s=\frac{\sigma}{\sigma_0}$, effectively oscillate with the
same frequency, approximately given by $\omega = \sqrt{2}$, around the
supersymmetric minimum, $\phi=0,\sigma=M/\sqrt{\lambda}$. For
energetic reasons the  phase difference amounts to $\pi/2$. Coming from the
critical point $(\phi=M/g,\sigma=0)$, depending on the initial values the dynamics in the
$\phi-\sigma$-plane should tend to be nearly one dimensional. Since in our
case the initial time derivates of the fields should be very small slow roll
values the trajectory should be near to a straight line.
Another interesting point, obvious without any calculation, is
that the dynamics of the post-inflationary system, measured in its natural
time and lengthscale $M^{-1}$, only depends on the scale of
the phase transition, $M/\sqrt{\lambda}$. As the M-rescaled classical fields $f$
and $s$ and their derivatives, take maximal values of ${\cal O} (1)$
 effectively, this scale has influence on
the damping but not on any other aspects of the background oscillations: the
smaller the scale is, the longer the system swings   correponding to its
natural timescale $M^{-1}$. Thus, qualitative
results of one breaking scale will also appear at another breaking scale.

This discussion  is confirmed by our numerical integration of the
system of equations, as shown in figures \ref{Linde0} and \ref{chaos}.
The parameters used in this and all of the following
investigations are given by
\begin{eqnarray}
\lambda&=&0.625\times 10^{-3}\nonumber\\
g^2&=&0.125\times 10^{-2}\nonumber\\
M&=&0.350\times 10^{15}\\
\frac{M}{\sqrt{\lambda}}&=&1.40\times 10^{16}\nonumber, 
\end{eqnarray}
which corresponds to a phase transition near to the SUSY-GUT scale in a
supersymmetric $SO(10)$ model in \cite{da}.
The solutions are essentially
damped oscillations around the sypersymmetric minimum, with the predicted
phase. The dynamics in the  $f$-$s$-plane is practically one
dimensonal and can be restricted to the straight line for very small initial values of the time derivatives of the fields. Anharmonic behaviour is caused by the interaction of the classical
fields and by the fact, that the Higgs field has a negative mass contribution
from the tachyonic mass.

\begin{figure}
\vspace*{0cm}
\begin{picture}(0,0)
\put(270,-30){ $\frac{R}{M^4/4\lambda}$}
\put(270,-132){ $\frac{R_{\phi}}{M^4/4\lambda}$}
\put(270,-85){ $\frac{R_{\sigma}}{M^4/4\lambda}$}
\put(80,0){\epsfysize=8cm\rotatebox{270}{\epsffile{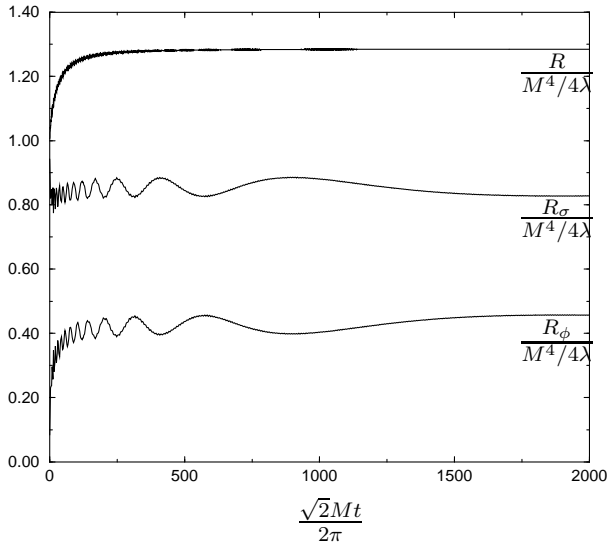}}}
\put(190,-200){$\frac{\sqrt{2}Mt}{2\pi}$}
\end{picture}
\vspace*{7cm}
\caption{energy densities of the background}\label{Energie}
\end{figure}

Our result for the supersymmetric case ($g^2=2\lambda$) is quite
different to the case with $g^2=\lambda$ , which was  considered in \cite{bellido2}, where
the classical fields oscillate in a rather chaotic way. But it essentially equals the
result in \cite{susypr}.

Now we consider the comoving energy densities after the phase transition:
\begin{eqnarray}
R&:=&\rho a^3,\\
R_{\phi}&:=&\rho_{\phi} a^3,\\
R_{\sigma}&:=&\rho_{\sigma} a^3
\end{eqnarray}
Here approximately $\rho = M^4/4{\lambda}$  is the system's total energy density at the
phase transition.
As the  Universe passes a phase of matter domination after inflation, such that $\rho\sim a^{-3}$, the
product $\rho a^3$  will be constant. This is what our numerical investigations
show  in figure \ref{Energie}. After the phase transition $R$  is growing for a short time, up to the value of $1.24M^4/4\lambda$
and then remains practically constant. It takes the background about 120
oscillations  until it swings  in such a  ,,harmonic'' way, that the universe behaves matter dominated.
Since we were interested in the effective distribution of energy between both
of the swinging classical fields, in the calculation of $R_{\phi}$ and
$R_{\sigma}$ we integrated out the short time dynamics ($t\sim {\cal
  O}(M^{-1})$) by averaging over the the background oscillations. Here we
respected the fact, that the frequency of the oscillations is slightly varying with time. The plot shows,
how energy is exchanged between the classical fields -  one reason for
anharmonic behaviour  in figure~\ref{Linde0}. After about 2000
oscillations the  energy distribution  between the oscillation of the inflaton
and Higgs-condensate is approximately $1/3:2/3$. Then the classical fields
essentially are decoupled oscillators.

\section{Parametric Excitation of Quantum Fluctuations}
\label{sec:4}
\setcounter{figure}{0}
\setcounter{equation}{0}
Our investigation of the quantum fluctuations uses the M-rescaling introduced in section \ref{sec:3}. We studied the mode equations following from the Heisenberg
expansion of a quantum field $\hat \chi$, which stands for the M-rescaled quantum
fluctuations $\delta\hat s(y,\vec \xi)$ and $\delta\hat f(y,\vec \xi)$ around the classical fields $s(y)$ and $f(y)$:
\begin{eqnarray}
\hat\chi (y,\vec x)&=& \frac{1}{(2\pi)^3}\int d^3p(\hat b_p X_p(y)e^{-i a\vec \xi \vec
  p}+\hat b^{+}_p X^*_p(y)e^{i  a\vec \xi \vec p}),\label{pEntw1}\nonumber\\
&=&\frac{1}{(2\pi )^3 a^{3/2}(y)}\int d^3K(\hat b_K X_K(y)e^{-i \vec \xi \vec
  K}+\hat b^{+}_K X^*_K(y)e^{i  \vec \xi \vec K}),\label{pEntw2}\nonumber\\
\end{eqnarray}
where $\vec p = \vec K/a$ is the M-rescaled physical momentum corresponding to
the M-rescaled commoving momentum $\vec K$.  
The operators $\hat b_p^{(+)}$
satisfy the commutation relation:
\begin{equation}
\left[\hat b_p,\hat b_{p'}^+\right]=\left( 2\pi\right)^3 \delta(\vec p - \vec p').
\end{equation}
Using the comoving momenta $K=|\vec K|=|\vec p|a$ for notation, this reads 
\begin{equation}
\frac{1}{a(y)^3}\left[\hat b_K,\hat b_{K'}^+\right]=\left( 2\pi\right)^3 \delta(\vec K - \vec K').
\end{equation}
Then the occupation number is given by
\begin{equation}\label{Beszahl}
n_K=\frac{\omega_K}{2}\Big(\frac{|\dot X _K|^2}{\omega _K ^2}+|X _K|^2\Big)-\frac{1}{2},
\end{equation}
and the particle density reads:
\begin{equation}\label{Teilchenzahl}
n_{\chi}(y)=\int \frac{d^3 K}{(2\pi a)^3} n_K.
\end{equation} 
Neglecting higher order terms in the quantum fluctuations, the mode equations read:
\begin{equation}
\left(\frac{d^2}{dy^2} +\left(\frac{K^2}{a^2}+2s^2-\frac{3}{2} \frac{\ddot a
  a+ \dot a^2/2}{a^2}\right) \right)
 {\delta f_K} (y)=0
\end{equation}
\begin{equation}
\left(\frac{d^2}{dy^2} +\left(\frac{K^2}{a^2}+3s^2+f^2-1-\frac{3}{2} \frac{\ddot a
  a+ \dot a^2/2}{a^2}\right) \right) 
{\delta s_K} (y)=0,
\end{equation}
As numerical calculations show, very soon $(3/2)(\ddot a a+\dot a^2/2)/(a^2)\sim 0$ is fullfilled within the numerical accuracy and this term can be neglected. This corresponds to a matter dominated universe, where typically $\ddot a a=-\dot a^2/2$ . We performed a numerical integration of the full set of differential equations. The values of $K$, for which these investigations were done, were
  selected by a pre-investigation using the well known approximation by the
  Mathieu equation \cite{kofman2}.

Before turning over to our numerical results, it is sensible first to
consider the problem of particle interpretation. As clearly pointed out
before, both of the classical fields, $\phi $ and $\sigma$,  are approximate representations
of the quantum fields $\hat \phi(t,\vec x)$ and $\hat \sigma(t, \vec x)$ in
the low momentum zone. Although we identified these classical fields with
their zero-momentum modes in principle they  have an extension in momentum
space. The edge of the condensates in momentum space will be situated at a momentum which is somewhat smaller than  the mass of
the corresponding field. In this sense the use of a homogeneous classical
field is similiar to a limitation of the ,,resolution'' in an optical investigation.
But if this is true, interpreting the quantum fluctuations around these
classical fields as free particles in presence of a homogeneous background  is justified only for
those modes, whose momenta are well above the edges of the condensates. 
This
important point was apparently not taken into consideration by the
authors of \cite{susypr}. 
\begin{figure}
\begin{minipage}[t]{6.5cm}
\vspace*{0cm}
\begin{picture}(0,0)
\put(0,-70){$\delta s_k$}
\put(10,0){\epsfysize=6.4cm\rotatebox{270}{\epsffile{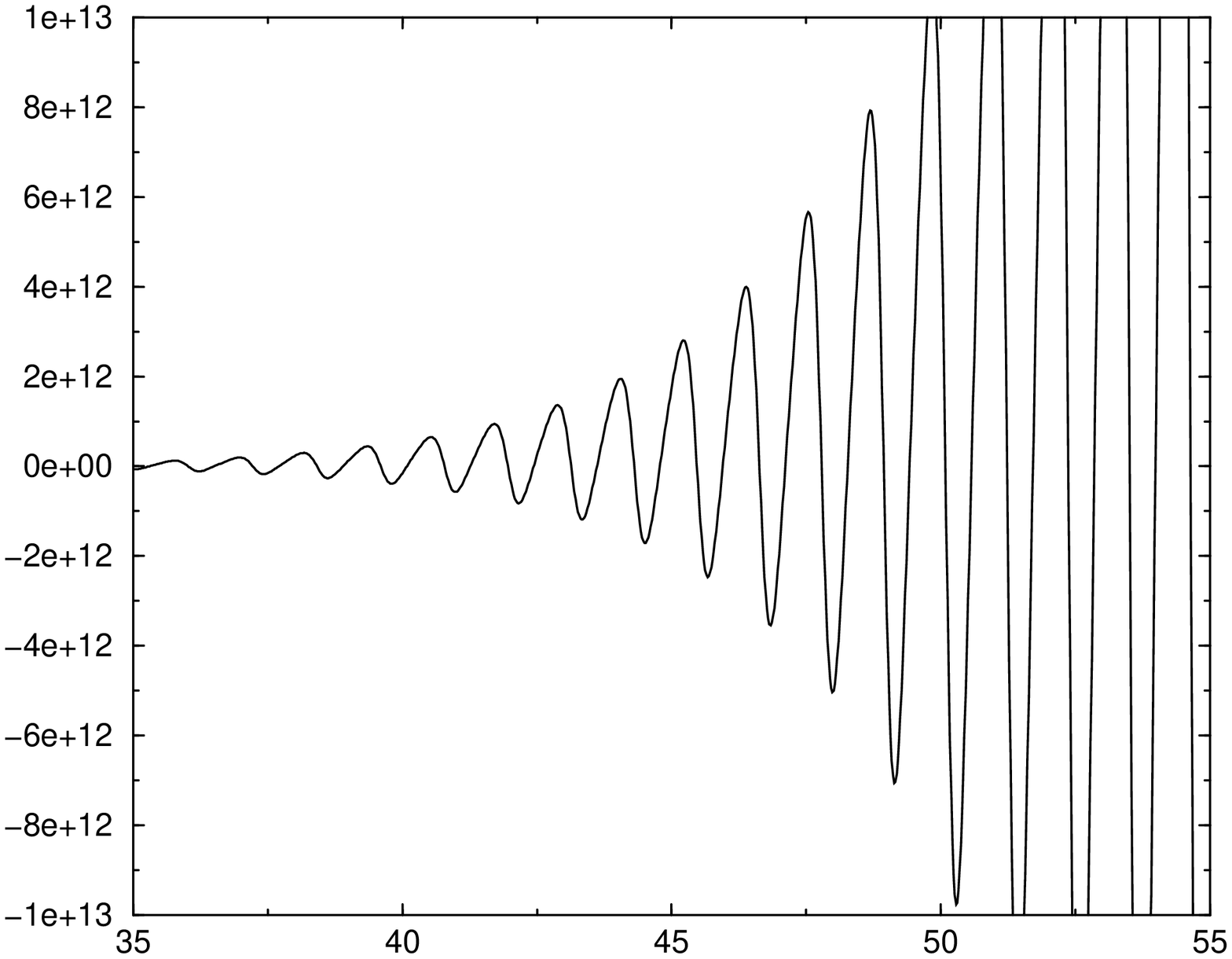}}}
\put(90,-152){$\frac{\sqrt{2}Mt}{2\pi}$}
\end{picture}
\vspace*{5cm}
\caption{mode function $\delta s_k$ for $k=M/10$}\label{testds}
\end{minipage}\hfill
\begin{minipage}[t]{6.5cm}
\begin{picture}(0,0)
\put(0,-70){$n_k$}
\put(10,0){\epsfysize=6cm\rotatebox{270}{\epsffile{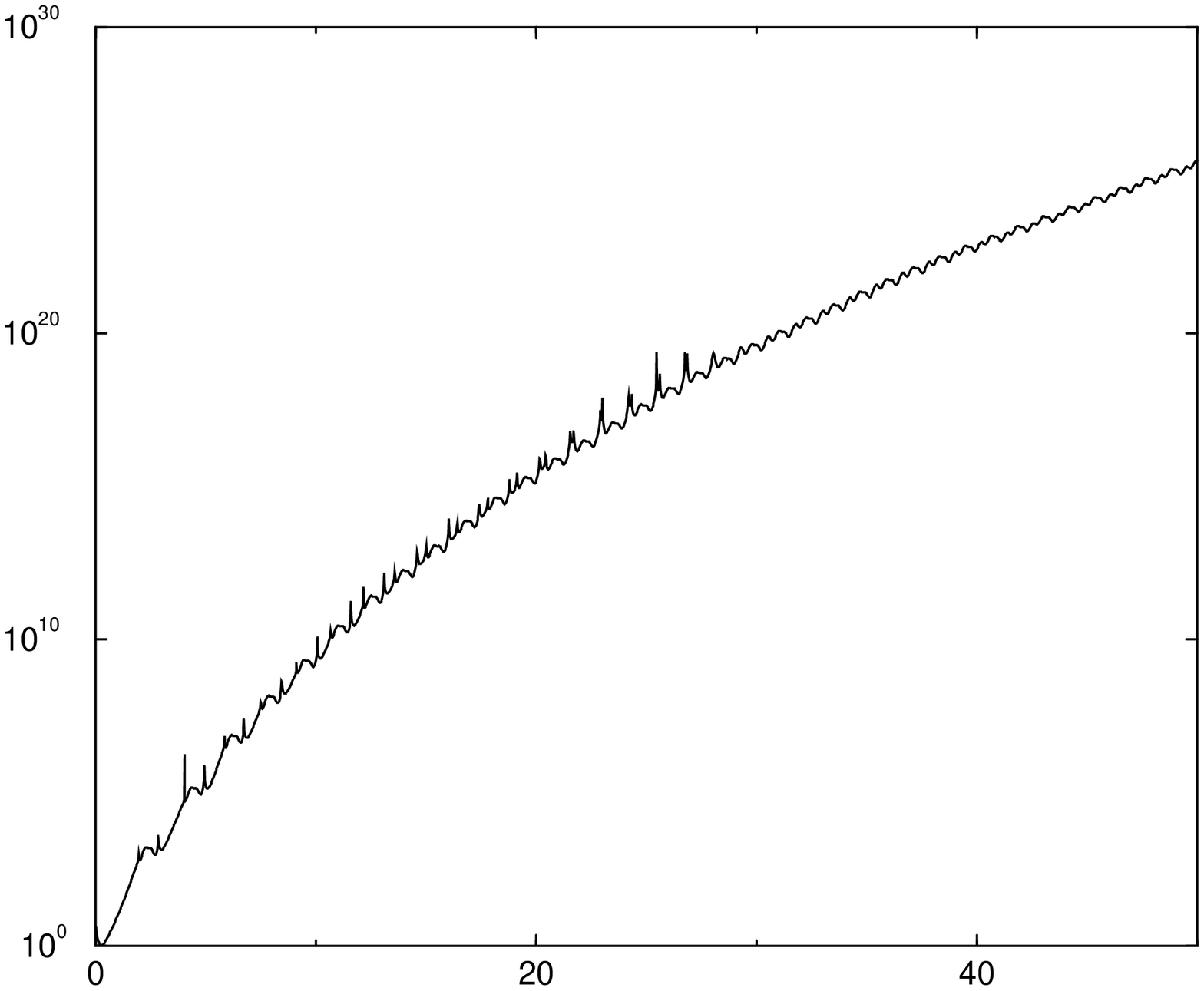}}}
\put(90,-152){$\frac{\sqrt{2}Mt}{2\pi}$}
\end{picture}
\label{fig2}
\vspace*{5cm}
\caption{resonance of the mode with $k=M/10$}\label{testns}
\end{minipage}
\end{figure}

As in the last section, we performed numerical calculations of the full system
of differential equations mentioned above. We did not find any parametric
amplification of the Higgs-fluctuations with $K\gsim 1/3$. For
$K\lsim 1/3$ there exists an amplification which can be described as a mixing of
both effects: mainly parametric resonance and partly tachyonic mass. In the well known
picture of the Mathieu equation,
\begin{equation}\label{Mathieugln}
\left( \frac{d^2}{dz^2}+A_k-2q\cos (2z) \right) \chi _k =0,
\end{equation}
which is applicable restricting  only to some few
oscillations, this resonance starts inside the first
resonance band ($1-q\lsim A_k\lsim 1+q$)  with a resonance parameter $q \sim 0.9$, which corresponds to
a strong narrow resonance regime. The amplification of
these modes is essentially independent of the momentum and is given by the solution
for $K=0$. Furthermore, these modes lie in a range of momenta, which we
estimated in section \ref{sec:2} to have non-vanishing fourier modes of the Higgs-condensate:
Because of this  we do not
consider these soft modes to be free particles. We think, that using the homogeneous classical field ansatz it does not make sense
to study Higgs fluctuations with such momenta.  Our result only
tells us someting about the extension of the distribution of $\hat \sigma$
which was described in section \ref{sec:2}.

The  picture in the case of inflaton fluctuations is similiar. Here a
tachyonic mass is missing and the effect of the resonance is much weaker. But
there is no amplification of modes with $K\gsim 1/10$ and similiar to the
above case,the observed
weak parametric amplification for modes with $K\lsim 1/10$ cannot be intepreted as   a creation of
free particles. 

In order to learn more about the space/momentum dependence of the condensates
one could  calculate the
mode equation of the fourier modes $s_K$ and $f_K$ of the
condensates. Since the original equations of motions are non-linear, the mode equations  then are  be rather complicated integro-differential equations. A first step of simplification
would be to let their masses only be given by  the zero momentum modes of the condensates. As can be checked easily, the  mode equations for $f_K$ are exactly the same as for $\delta f_K$. 
Those for $s_K$ will look a little bit different from those for $\delta s_k$ but will have
similiar resonance parameters in the Mathieu approximation. Therefore, from
the investigation that we did for the quantum fluctuations one can conclude
that there will be a similiar resonance effect in a similiar range of momenta
for the modes $s_K$ and $f_K$.  

This shows that after the phase transition the inflaton
condensate slightly looses its homogenity, while the Higgs condensate is
non-homogeneous from the very beginning. However, at a physical scale which is somewhat
bigger than $aM^{-1} $ they are
essentially homogeneous. Quantum fluctuations around these condensates should have monumenta which are bigger than the inverse scale of homogenity, in order to have a particle interpretation.

Until now our investigations did not consider the effect of
backreaction of the quantum fluctuations on
the classical fields, which is very important in the case of a preheating with
a very efficient particle production. As pointed out in \cite{kofman2}, the main effect of the backreaction is
to lower the resonance parameters in the picture of the Mathieu
equation. This means, that appearing resonances will become weaker and will last for
a shorter time. Then, in our case, the general picture is not changed. The
extension of the condensates in momentum space probably will be smaller, such that they will be  even  ``more homogeneous'' than without
backreaction. But there will be no possibility of having a parametric
amplification of fluctuations, which were not amplified also without backreaction.

So we
conclude that there does not exist a creation of free Higgs- or singlet-particles
by parametric amplification. This corresponds to the result in \cite{bellido2} for
the different case of $\lambda=g^2$.

\section{Parametric Excitations of External Fields}
\label{sec:5}
Although we did not find a  preheating of the quantum fluctuations, there could be a parametric creation of an external scalar quantum
field $\hat \chi$, that couples to the Higgs sector via
\begin{equation}
\frac{1}{2}h_1^2\phi^2\hat\chi^2+\frac{1}{2}h_{2}^2\sigma^2\hat\chi^2.
\end{equation}
Then the equations of motions will be modified in an obvious way. The
mode equation for the modes $\chi_K(t)$  reads
\begin{equation}
\left(\frac{d^2}{dy^2} +\left(\frac{K^2}{a^2(t)}+\frac{h_1^2}{g^2}f^2+\frac{h_2^2}{\lambda}s^2\right) \right) 
\chi_K (y,\vec \xi)=0,
\end{equation}
where again $\ddot a a =-\dot a^2/2$ was assumed.
\setcounter{figure}{0}
\setcounter{equation}{0}
\begin{figure}
\vspace*{0cm}
\begin{picture}(0,0)
\put(70,-85){$n_k$}
\put(80,0){\epsfysize=8cm\rotatebox{270}{\epsffile{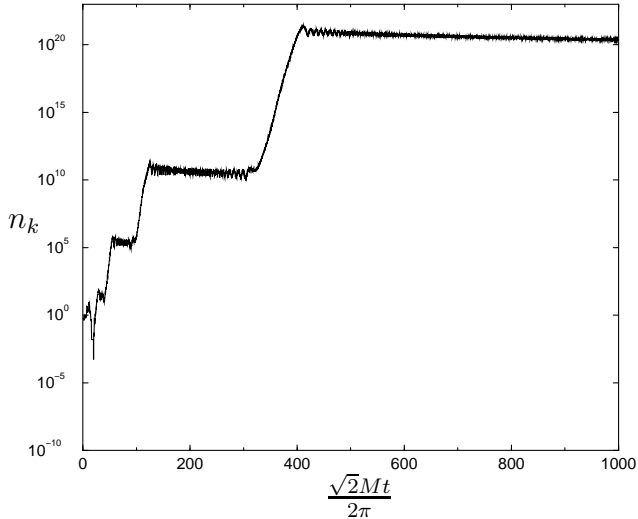}}}
\put(190,-190){$\frac{\sqrt{2}Mt}{2\pi}$}
\end{picture}
\vspace*{6.5cm}
\caption{parametric excitation of the $k=M$-mode of a scalar field $\hat \chi$}\label{nk}
\end{figure}

We used the same methods as before in our numerical investigation. Our result
is very similiar  to the result in \cite{bellido2} for $g^2=\lambda$. We found a strong resonance only for the case
without coupling to the Higgs-condensate, $h_2\simeq 0$ (see figure \ref{nk}). But this situation is
impossible, if we use the superpotential (\ref{RWPot}). The $F$-terms always lead to
$h_2\not= 0$ if $h_1\not= 0$ and vice versa. As can be shown very easily \cite{da},
in this model
there does not exist any coupling  allowed by 
supersymmetry and  leading to a bosonic parametric resonance. This result, as far as we can see, 
buries the hope for the existence of a bosonic preheating.

\section{Consequences for Left-Right Symmetric Models}
In left-right-symmetric models the phase transition that leads to  the gauge symmetry of
the Standard Model, necessarily lowers the rank of the Lie group. The only renormalizable superpotential with
this feature has the structure
\begin{equation}
  \label{eq:bla}
  W=X\bar C C+ \textrm{polynomial in }X,
\end{equation}
where $X$ is a gauge singlet and $\bar C, C$ are appropriate spinor representations of the
left-right-symmetric gauge group. It is an interesting question, if, using the
superpotential (\ref{RWPot}), there could be a natural  embedding of inflation
into the context of a right-left-symmetric model such as $SO(10)$ \cite{da}
or $SU(3)_c\times SU(2)_L\times SU(2)_R\times U(1)_{(B-L)}$, which
leads to a senseful cosmological model.  

In minimal $SO(10)$-models \cite{babu2,barr} for example $C$  should be  a
16-dimensional spinor representation, the scalar component of which aquires a
GUT-scale vev in the $SU(5)$-singlet direction during the phase
transition. Since the $SO(10)$ (more precisely: $Spin(10)$) is a simple and
simply connected Lie group, breaking it down to the SM-gauge group will lead
to unwanted monopoles. They appear in a first phase transition, when a 45
dimensional Higgs representation  aquires a suitable vev  to break $Spin(10)$
down to the left-right-symmetric group  $G_{LR}:=SU(3)_c\times SU(2)_L\times
SU(2)_R\times U(1)_{(B-L)}$. The inflaton then dilutes the unwanted remnants
and ends in a phase transition which   breakes the remaining gauge symmetry to
the standard model group. This picture remains the same in a pure
$G_{LR}$-Model, $C$ being a $SU(2)_R\times U(1)_{(B-L)}$-Higgs-doublet.

Our investigations indicate, that in such models 
free bosonic particles cannot be created by the effect of parametric resonance
after inflation. The parametric amplification in the ,,low momentum zone'' of
the quantum fluctuations, which appears in our numerical investigation, cannot be interpreted as a production of free particles. In this
area not only the dynamics of the mode functions are essentially independent of the momentum but also the
validity of the homogeneous classical field ansatz is doubtful. In our view,
these numerical results perhaps can tell us something about the extension
of the condensates in momentum space but shurely show us the limitation of the
homogeneous classical field ansatz.

Supposing that bosonic preheating does not take place, a next step would be to
study the possibility of
fermionic preheating for this superpotential, which recently was shown to be possibly very efficient \cite{ferm}. Since we deal with a supersymmetric theory, there is no
reason for neglecting  the fermionic superpartners. Also a parametric
creation of Majorana neutrinos could appear, which should be very interesting
in the context of leptogenesis. In this context it also would be nessecary to
rule out the creation  of the helicity-1/2-gravitinos  by non-perturbative
effects, which recently was
shown to be of possible danger for cosmological models which  involve
supersymmetry \cite{grav}. 
If fermionic preheating turned out not to be efficient, particle production in left-right-symmetric models could be caused only by
perturbative effects. Following \cite{reh} then still the production of super heavy
matter could be 
possible. But, as we work with two condensates, both of them would have to
decay very efficiently, in order to be not in conflict with standard cosmology. Considering the gravitino constraint on the reheating
temperature of the Universe, this would mean strong limitations to possible
couplings to other fields. Concluding, there is still some work to do, until the particle
creation after inflation in this very simple supersymmetric model will be fully investigated. 

\section*{Acknowledgment}
We would like to thank S.~J.~Huber and P.~John for 
useful discussions.
This work was supported in part by the
TMR network {\it Finite Temperature Phase Transitions in Particle 
Physics}, EU contract no. ERBFMRXCT97-0122.

\include{elitpaper}

\end{document}